# A computational model of radiolytic oxygen depletion during FLASH irradiation and its effect on the oxygen enhancement ratio

Guillem Pratx, PhD* and Daniel S Kapp, MD, PhD
Department of Radiation Oncology, Stanford University School of Medicine

* Corresponding author
pratx@stanford.edu
(650) 724 9829
300 Pasteur Dr, Grant S277
Stanford, CA 94305-5132

**Abstract**
Recent results from animal irradiation studies have rekindled interest in the potential of ultra-high dose rate irradiation (also known as FLASH) for reducing normal tissue toxicity. However, despite mounting evidence of a "FLASH effect", a mechanism has yet to be elucidated. This article hypothesizes that the radioprotecting effect of FLASH irradiation could be due to the specific sparing of hypoxic stem cell niches, which have been identified in several organs including the bone marrow and the brain. To explore this hypothesis, a new computational model is presented that frames transient radiolytic oxygen depletion (ROD) during FLASH irradiation in terms of its effect on the oxygen enhancement ratio (OER). The model takes into consideration oxygen diffusion through the tissue, its consumption by metabolic cells, and its radiolytic depletion to estimate the relative decrease in radiosensitivity of cells receiving FLASH irradiation. Based on this model, several predictions are made that could be tested in future experiments: (1) the FLASH effect should gradually disappear as the radiation pulse duration is increased from <1s to 10 s; (2) dose should be deposited using the smallest number of radiation pulses to achieve the greatest FLASH effect; (3) a FLASH effect should only be observed in cells that are already hypoxic at the time of irradiation; and (4) changes in capillary oxygen tension (increase or decrease) should diminish the FLASH effect.

# I. Introduction

For decades, radiobiological studies have sought to understand the acute and late effects of dose rate of ionizing radiation in attempts to optimize the therapeutic ratio of tumor therapy. Early studies demonstrated a "breakpoint" in the clonogenic survival curves of both bacterial and mammalian cells irradiated at ultra-fast dose rates (>10 Gy/s), with doses above the breakpoint dose resulting in clonogenic cell survival transitioning from aerobic to anaerobic survival (Berry and Stedeford, 1972; Epp *et al.*, 1972; Nias *et al.*, 1969). The breakpoint dose was shown to be consistent with radiolytic oxygen depletion (ROD), a transient decrease in oxygen tension caused by radiochemical reactions (Weiss *et al.*, 1974). However, since the effect was seen only in cells already at reduced levels of oxygen (<4 mmHg), it was concluded by some that "no special advantages are likely to accrue from the use of radiations at ultra-high dose rates for the radiotherapy of human cancers" (Berry and Stedeford, 1972).

More recently, investigators at the Institut Curie in France and the Centre Hospitalier Universitaire Vaudois in Switzerland have reported unexpectedly low normal tissue toxicities in animals receiving ultra-high dose rate irradiation. Using short pulses (<1 s) of radiation delivered at dose rates of 40 Gy/sec or greater (known as FLASH), they analyzed the response of normal animal tissues to radiation and found less pulmonary fibrosis and less epithelial cell acute radiation apoptosis (Favaudon *et al.*, 2014), sparing of memory (Montay-Gruel *et al.*, 2017), and reduced depilation and fibronecrosis (Vozenin *et al.*, 2019a). In addition, a FLASH effect has also been recently reported on the induction of radiation-induced alteration of zebrafish morphology following irradiation of fertilized zebrafish eggs (Vozenin *et al.*, 2019b). The potential of FLASH irradiation for improving the therapeutic ratio has sparked significant interest in the radiation oncology community and the topic has been cover by several recent review articles (Durante *et al.*, 2018; Vozenin *et al.*, 2019b; Al-Hallaq *et al.*, 2019).

However, despite mounting evidence of normal tissue sparing by FLASH radiation therapy, a mechanism of action has not yet been demonstrated (Durante *et al.*, 2018). This study examines whether ROD of normal tissue stem cells in hypoxic niches could explain the radioprotection of normal tissues during FLASH irradiation. Normal tissues are known to be well oxygenated and, based on existing *in vitro* evidence, no biologically significant ROD would be expected in these tissues at clinical dose levels (i.e., 10 Gy). However, normal tissue oxygenation is also highly heterogeneous, and the effect of FLASH irradiation may vary accordingly. We thus hypothesize that the radioprotecting effect of FLASH irradiation could be through the specific sparing of hypoxic stem cell niches, which have been highlighted in several organs including the bone marrow and the brain. Due to the lower oxygenation of these niches, increased sparing of normal-tissue stem cells may be achieved through FLASH-mediated ROD. Stem cells are known for their powerful tissue regeneration capabilities. For instance, in a previous study, investigators have shown that the transfer of $4\times10^5$ unirradiated stem cells to the hippocampus of brain-irradiated mice was sufficient to rescue mice from significant cognitive impairments (Acharya *et al.*, 2009).

To further investigate the role of hypoxic stem cell niches in FLASH irradiation, we perform computational simulations of ROD according to known parameters for oxygen depletion rate, oxygen concentration and oxygen diffusion. We assess the effect of FLASH irradiation by considering the relative decrease in the oxygen enhancement ratio (OER) due to ROD, a factor we call $\delta_{ROD}$. Following this analysis, we make several predictions regarding the effect of FLASH radiotherapy for different radiation doses, dose rates, oxygen levels, and pulse structures. Testing of these predictions using experimental models could be useful to help elucidate the mechanism of FLASH irradiation.

## II. ROD at ultra-high dose rate (>100 Gy/s)

*II.A. Theory*

We first consider the case where radiation is delivered fast enough that the rate of ROD greatly exceeds changes in oxygen tension due to oxygen diffusion and metabolism in the tissue. In this first model, oxygen diffusion and metabolism are assumed to be negligible. A second model is presented in the next section where these effects are taken into consideration.

The effect of oxygen (or lack thereof) on the response to ionizing radiations has been extensively studied in vitro, in preclinical models, and in patients (Vordermark and Horsman, 2016). This effect is typically quantified according to the oxygen enhancement ratio (OER), the ratio of the dose in anoxia to the dose in air required to achieve a defined rate of cell survival, and has been measured under a wide range of radiation dose rates (Ling et al., 1985; Ling et al., 2010). Oxygen enhancement requires oxygen to be present at the time of the irradiation (or a few microseconds after irradiation) and is due to two factors, (*i*) oxygen-enhanced formation of deleterious free radicals and (*ii*) rapid fixation of DNA damage (the so-called oxygen fixation hypothesis (Schwartz, 1952; Alper and Howard-Flanders, 1956a)). Several parameterizations have been proposed to fit experimental cell survival data according to oxygen level. For instance, the classical Alper-Howard-Flanders equation models the OER as a ratio of two parameters (Alper and Howard-Flanders, 1956a). A more recent parameterization is that of Grimes and Partridge (Grimes and Partridge, 2015), used in this study,

$$\text{OER}(p) = 1 + \frac{\Phi_0}{\Phi_D}(1 - e^{-\varphi p}) \qquad (1)$$

where $\Phi_0/\Phi_D = 1.63$ and $\varphi = 0.26$.

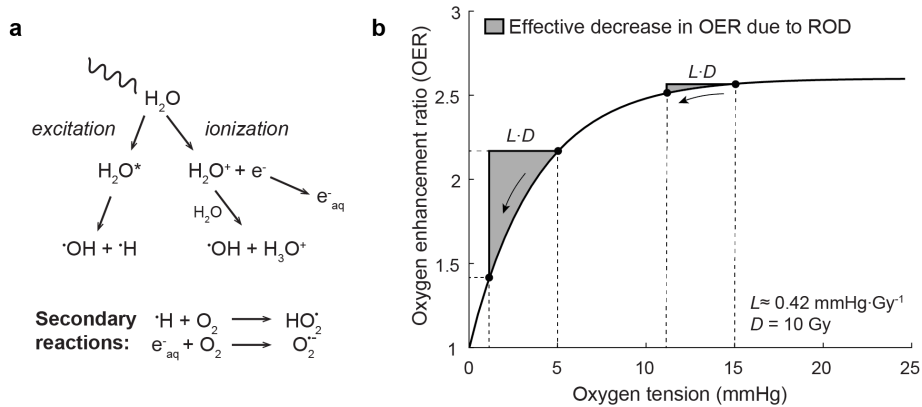

**Figure 1. Radiolytic oxygen depletion (ROD) and its effect on the oxygen enhancement ratio (OER).** (a) Water radiolysis results in formation of radicals, such as the $^\bullet$H and $e^-_{aq}$, that react with oxygen to yield superoxide ($O_2^{\bullet-}$) and its protonated form ($HO_2^\bullet$). (b) OER as a function of oxygen tension. The amount of oxygen depleted by a single-fraction dose $D = 10$ Gy is $L \cdot D \approx 4.2$ mmHg (Weiss et al., 1974). Oxygen depletion occurs progressively during the delivery of the dose and so does the decrease in OER. The shaded area represents the effective decrease in OER integrated over the duration of the dose deposition. The OER curve is drawn according to Equation (1).

Radiochemistry experiments have shown that oxygen can be depleted during irradiation due to its reaction with two byproducts of water radiolysis, the solvated electron and the hydrogen radical (Figure 1a). Irradiation of water with low linear-energy-transfer (LET) radiation produces solvated electrons and hydrogen radicals at a combined rate of 5.1 molecules per 100 eV, equivalent to 0.5 µM/Gy (Colliaux et al., 2015). A water solution in equilibrium with 3.8 mmHg (0.5%) oxygen contains approximately 5.2 µM of dissolved oxygen according to Henry's law ($K = 960$ atm L/mol at 37°C), thus a radiation dose of 10 Gy would generate a sufficient number of oxygen-depleting radicals to achieve stoichiometry with oxygen in a reaction that would yield the superoxide radical and its protonated form, $HO_2^\bullet$ (Figure 1a). While cytotoxic, these radicals can also be neutralized by enzymes within cells. What is more important, however, is that oxygen depleted through this radiolytic process is no longer available to fix DNA damage, leading to a potential increase in clonogenic cell survival.

The depletion of oxygen through water radiolysis has also been reported experimentally in hermetically sealed solutions. Oxygen is depleted from these solutions in proportion to the radiation dose applied, at rate $L = 0.21$-$0.42$ mmHg/Gy (Whillans and Rauth, 1980; Weiss et al., 1974; Michaels, 1986). This rate is independent of initial oxygen concentration. Ignoring oxygen diffusion and metabolism, the oxygen tension following a FLASH dose of radiation $D$ is

$$p = p_0 - LD, \qquad (2)$$

where $p_0$ is the oxygen tension before irradiation.

Accordingly, the effect of ROD can be qualitatively understood as a shift of the OER curve by an amount equal to $LD$. However, it is also important to consider that ROD occurs on a similar time scale as DNA damage fixation (Colliaux et al., 2015). Starting from a baseline level $p_0$, oxygen is gradually depleted through ROD according to Equation (2), where $D$ varies from 0 to the prescribed dose $D_p$. As this process takes place, the tissue is made progressively more resistant to radiation due to the continuous decrease in OER.

Considering an infinitesimal dose increment $dD$, the fractional cell kill can be expressed as:

$$\frac{dN}{N(D)} = -\alpha(p_0 - LD)dD, \qquad (3)$$

where $N(D)$ is the number of surviving cells after a dose $D$ and $\alpha(p)$ is the radiosensitivity under oxygen tension $p$. Since $LD$ can exceed the $p_0$, the radiosensitivity function is extended such that $\alpha(p) = \alpha(0)$ for $p < 0$ (complete oxygen depletion).

Next, the fraction of cells killed is integrated up to prescribed dose $D_p$ to yield the fractional cell survival

$$\log\left(\frac{N_p}{N_0}\right) = -\int_0^{D_p} \alpha(p_0 - LD)\, dD \qquad (4)$$

where $N_p \equiv N(D_p)$ is the number of surviving cells after prescribed dose $D_p$. This formula can be applied to the conventional irradiation by setting $L = 0$ (no ROD), resulting in the classical exponential cell kill model

$$\log\left(\frac{N_p}{N_0}\right)\bigg|_{L=0} = -\alpha(p_0)D_p. \qquad (5)$$

To better understand the effect of ROD during FLASH irradiation relative to conventional irradiation ($L = 0$), we reformulate Equation (4) as

$$\log\left(\frac{N_p}{N_0}\right) = (1-\delta_{ROD})\log\left(\frac{N_p}{N_0}\right)\bigg|_{L=0}. \qquad (6)$$

where $\delta_{ROD}$ is defined as

$$\delta_{ROD} = \frac{\alpha(p_0)D_p - \int_0^{D_p}\alpha(p_0-LD)dD}{\alpha(p_0)D_p}. \qquad (7)$$

This term is a function of two factors, the oxygen tension $p_0$ and the prescribed dose $D_p$. Because a linear radiosensitivity model is assumed, $\delta_{ROD}$ can also be expressed as a function of the OER:

$$\delta_{ROD} = \frac{OER(p_0)D_p - \int_0^{D_p} OER(p_0-LD)dD}{OER(p_0)D_p}, \qquad (8)$$

where OER is defined as a ratio of radiosensitivities

$$OER(p) = \frac{\alpha(p)}{\alpha(0)}. \qquad (9)$$

In the following, we use $\delta_{ROD}$ to quantify the relative effect of ROD during FLASH irradiation on fractional cell survival. For conventional irradiation, ROD is negligible and, therefore, $\delta_{ROD} = 0$. More generally, the term $\delta_{ROD}$ can be interpreted as the relative decrease in radiosensitivity attributable to ROD.

*II.B. Results*

Using the formalism developed in Section II.A, we investigate how $\delta_{ROD}$ varies as a function of the prescribed dose (single fraction) and the partial pressure of oxygen in the tissue.

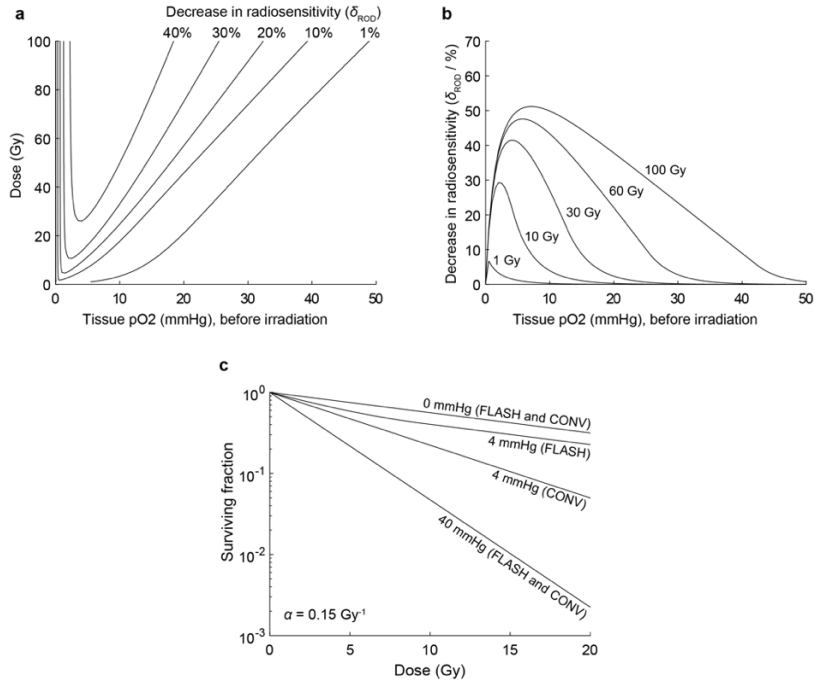

**Figure 2.** Relative decrease in radiosensitivity due to ROD for ultra-high-dose-rate radiation (>100 Gy/s). (a) The factor $\delta_{ROD}$ represents the relative decrease in the radiosensitivity of the tissue attributed to ROD, for different dose and oxygen levels. Black lines represent iso-effect lines. This model does not include oxygen

diffusion and metabolism, which occur on a slower time scale than ROD. (b) Same data, plotted for five different radiation doses. (c) Predicted cell survival curves plotted for different radiation delivery conditions. These curves assume a linear radiosensitivity model ($\alpha = 0.15$ Gy$^{-1}$).

We first plot $\delta_{\text{ROD}}$ according to various isoeffect lines, such as 1%, 10%, 20%, and so forth (Figure 2a). As the amount of dose delivered to the tissue increases, oxygen is increasingly depleted and cells become more radioresistant. This is better seen in Figure 2b, where we have plotted the same data for 5 different radiation doses. At the low-dose level (1 Gy), the effect of ROD on effective dose is marginal and limited to very low oxygen levels. For a dose of 10 Gy, we observe a significant decrease in radiosensitivity of up to 30% due to ROD. However, radioprotection is only imparted to cells residing within a narrow range of oxygen levels of 3-5 mmHg. For higher radiation doses (≥30 Gy), this range broadens considerably but these very high single-fraction doses are rarely used in the clinic.

We finally consider the effect of ROD on clonogemic cell survival at three different oxygen level, 0, 4 and 40 mmHg. Similar to Equation (3), radiosensitivity is assumed to be linear with dose. Under this assumption, we make the following observations: for anoxic or normoxic cells (40 mmHg), there is no difference between FLASH and conventional dose-rate radiation. It is only for hypoxic cells (pO$_2$ = 4 mmHg) that the two curves diverge (Figure 3c). The survival curve obtained during FLASH irradiation is initially parallel to the curve obtained for the conventional treatment. Once oxygen is fully depleted (which occurs at dose $D_p = \frac{p_0}{L} = 9.5$ Gy), the curve becomes parallel to the survival curve of anoxic cells.

### III. ROD at very-high dose rate (>10 Gy/s)

*III.A. Theory*

In the previous section, we assumed that changes in oxygenation due to cell respiration and tissue diffusion were negligible in comparison to ROD due to the high dose rate. Here, we develop a spatiotemporal model that includes the effect of these two important factors. The diffusion of oxygen from a capillary into the surrounding tissue is modeled according to the diffusion equation in polar coordinates. Here, we assume an infinitely long capillary (radius r$_0$ = 3 µm) and set oxygen tension within it to a fixed value $p_0$. Due to the equilibrium between dissolved oxygen and hemoglobin-bound oxygen, we can assume that ROD does not affect oxygen tension inside the capillary. We also assume a diffusion constant $D_{O2} = 2 \cdot 10^{-5}$ cm$^2$ s$^{-1}$ (Ferrell and Himmelblau, 1967) and a rate $m$ of oxygen consumption through cellular metabolism. As explained in the previous section, the rate of oxygen depletion is equal to $LD_p/T$, where $L$ is the ROD rate constant, $D_p$ is the prescribed dose, and $T$ is the radiation pulse duration. The diffusion and metabolism of oxygen in the tissue is described by the following equation:

$$\frac{\partial p}{\partial t} = D_{O2} \frac{1}{r}\frac{\partial}{\partial r}\left(r\frac{\partial p}{\partial r}\right) - m - \frac{LD_p}{T}. \tag{10}$$

We first calculate the steady-state oxygen tension around the capillary before irradiation ($D_p = 0$), which we call $p_{ss}(r)$. We set the boundary condition as $p_{ss}(r \leq r_0) = p_0$. The rate of oxygen metabolism $m$ is taken as a parameter of the model. Based on these curves, we select a metabolic rate of 3 mmHg/s, resulting in diffusion of oxygen up to 100 µm from the capillary (Figure 3).

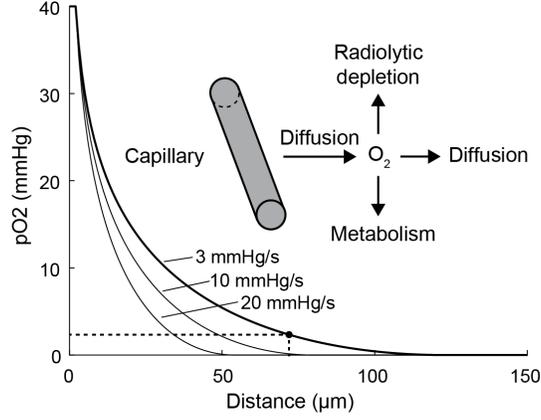

**Figure 3. Oxygen diffusion and metabolism during FLASH irradiation.** The model assumes a single infinite capillary (radius 3 µm, pO₂ = 40 mmHg) surrounded by metabolic cells. Oxygen diffusing from the capillary is consumed either by cells or through ROD. Oxygen concentration in steady-state (no radiation applied) is plotted as a function of the distance from the capillary for three levels of cell metabolism.

The diffusion equation is then solved again to model transient oxygen depletion during FLASH irradiation. The boundary conditions for this equation are $p(t = 0, r) = p_{SS}(r)$ and $p(t, r \leq r_0) = p_0$. Oxygen tension in the tissue $p(t, r)$ is then computed as a function of time and distance from capillary using the finite difference method ($\Delta r = 2$ µm and $\Delta t = 0.1$ µs) implemented in MATLAB (version R2015b).

Once the oxygen tension $p(r, t)$ is known, the survival cell fraction using the approach described in the previous section:

$$\log\left(\frac{N_p}{N_0}\right) = -\frac{D_p}{T}\int_0^T \alpha(p(r,t))\, dt \qquad (11)$$

Here, the integration is performed over time, which is equivalent to Equation (4) given that $D = \frac{LD_p}{T}t$. Similar in Equation (6), we introduce $\delta_{ROD}$ to quantify the decrease in radiosensitivity relative to conventional irradiation (no ROD, $L = 0$):

$$\delta_{ROD} = \frac{OER(p_{SS}) - \frac{1}{T}\int_0^T OER(p(r,t))\, dt}{OER(p_{SS})}. \qquad (12)$$

It should be noted that, according this definition, $\delta_{ROD}$ is a complicated factor that depends on several parameters, including oxygen pressure in the capillary ($p_0$), duration of the radiation pulse ($T$), total dose delivered ($D_p$), and distance to the capillary ($r$).

*III.B. Results*

At very high dose rates (>10 Gy/s), ROD remains a significant effect, yet oxygen diffusion and metabolism also affect the balance of oxygen within the tissue. Using the diffusion equation, we first simulate transient oxygen depletion during irradiation with 10 Gy pulses of different durations (Figure 4a). To simplify our analysis of the results, we focus on a small volume of tissue located 75 µm from a capillary, thus hypoxic (pO₂ = 2.1 mmHg). First, we consider a single 10 µs pulse of 10 Gy (dose rate = $10^6$ Gy/s). At this high dose rate, the rate of ROD greatly exceeds oxygen diffusion and no significant replenishment occurs during the delivery of the pulse, and oxygen is entirely depleted. Full replenishment of the irradiated tissue occurs 2-3 seconds after irradiation. We then simulate a longer pulse of 1 s (dose rate = 10 Gy/s). Due to the lower dose rate, oxygen lost through ROD is partially replenished, preventing its complete depletion. Finally, longer pulses of radiation (10 and 100 s) are unable to deplete oxygen from the tissues as required for significant radioprotection.

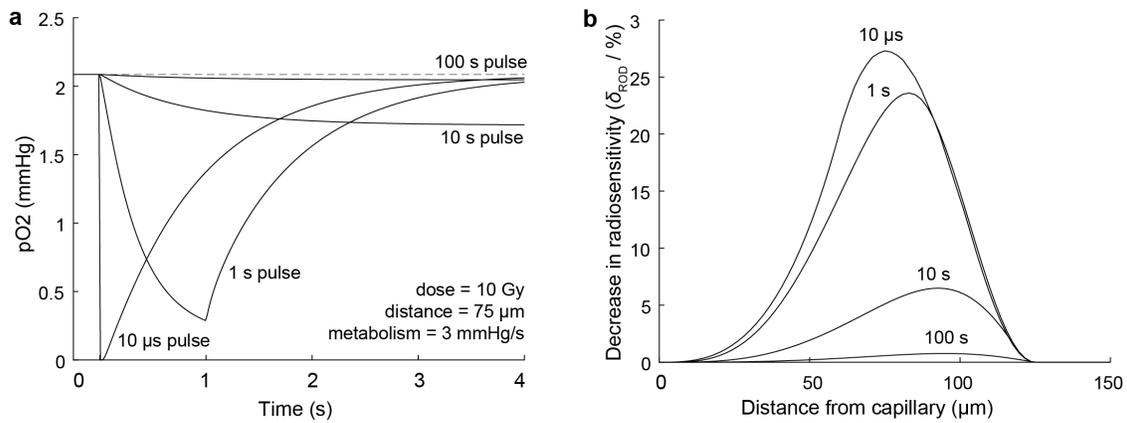

**Figure 4. Effect of ROD for different pulse durations.** (a) Transient decrease in oxygenation due to ROD during irradiation by 10 Gy pulses of various durations. The pulse is initiated at $t = 0.1$ s and the oxygen level is observed 75 μm from a capillary containing 40 mmHg $O_2$. (b) Effect of transient oxygen depletion on radiosensitivity, as quantified by the factor $\delta_{ROD}$, as a function of the distance to capillary.

We then consider the anticipated effect of pulse duration on radiosensitivity (as measured by $\delta_{ROD}$) as a function of the distance to the capillary. For the ultra-short pulse (10 μs), the strongest decrease in radiosensitivity is achieved in tissues 60-100 μm from the capillary (Figure 4b). Based on radiosensitivity, these tissues would experience an effective decrease in cell radiosensitivity of 20-25% compared to conventional dose rate irradiation. For the 1 s pulse duration, a similar albeit weaker effect is observed due to partial reoxygenation during dose delivery. Finally, only a marginal decrease in cell radiosensitivity (<5%) is expected for longer pulses (10 and 100 s).

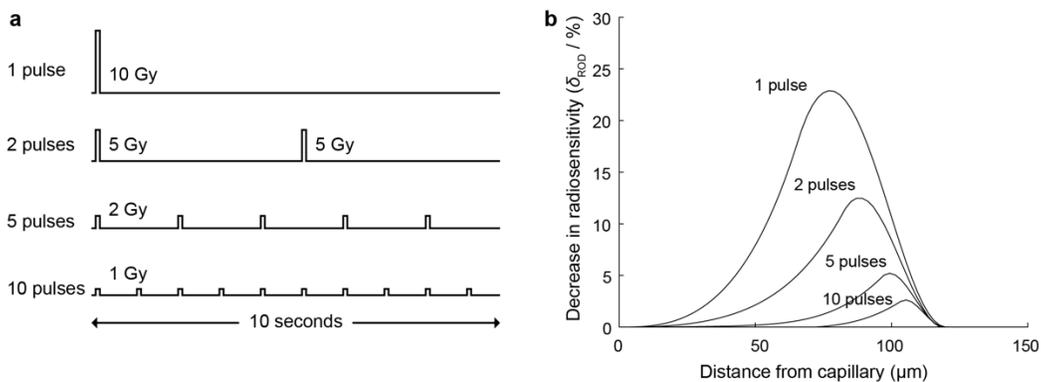

**Figure 5. Effect of ROD for different pulse structures.** (a) A single-fraction dose of 10 Gy is delivered as a series of 10 μs pulses. (b) The resulting effect is evaluated by computing $\delta_{ROD}$ for different pulse structures.

We then turn to the effect of pulse structure on ROD and radiosensitivity. In this simulation, a total dose of 10 Gy is delivered over 10 seconds as a series of discrete 10 µs pulses (Figure 5a). Clinical linear accelerators deliver dose as 5 µs pulses, typically at a rate of 60-360 Hz. Results from these simulation shows that the effect of ROD on radiosensitivity becomes negligible as the pulse repetition rate is increased (Figure 5b). This is because oxygen is replenished between each pulse, limiting its depletion. The strongest effect on $\delta_{ROD}$ was measured for single pulses. Fractionation of the radiation into 10 pulses (each separated by 1 sec) caused $\delta_{ROD}$ to drop below 5%. As radiation was fractionated into an even greater number of pulses, the effect on ROD became undistinguishable from continuous dose delivery.

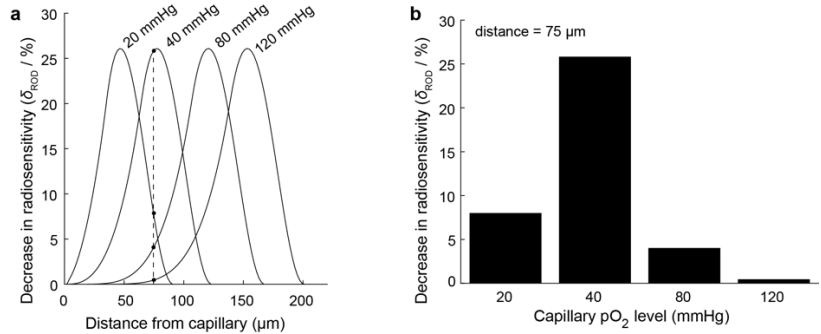

**Figure 6. Effect of ROD for different capillary oxygen levels.** (a) Relative decrease in radiosensitivity due to ROD ($D_p$ = 10 Gy, $T$ = 10 µs) shown for different capillary oxygen tensions. (b) Considering a volume of tissue 75 µm away from the capillary, the effect of ROD on radiosensitivity ($\delta_{ROD}$) depends substantially on oxygen tension in the capillary.

Finally, we determine how changes in capillary oxygen tension may increase or decrease the magnitude of ROD during FLASH irradiation. Tissue oxygenation could be manipulated experimentally by carbogen and nitrogen breathing to help better understand the potential role of ROD to normal tissue sparing by FLASH radiotherapy.

According to the model, changing oxygen tension in the capillary shifts the location in the tissue where ROD yields the greatest radioprotecting effect but does not affect the magnitude of the effect (Figure 6a). For instance, if capillary oxygen tension is decreased from 40 to 20 mmHg, the region in which ROD protects cells will move closer to the capillary, from 75 µm to 45 µm, but the decrease in radiosensitivity $\delta_{ROD}$ will remain the same. Conversely, if oxygen levels are increased, this region will move further away from the blood supply. This effect is shown in Figure 6b, where the effect of ROD on radiosensitivity is shown for a region of tissue distant 75 µm from the capillary. Overall, these data suggest that radioprotection of cells by ROD would be strongly affected by changes in capillary oxygenation.

## IV. Discussion

We presented two different models of ROD during FLASH irradiation. In the first model, we consider radiation delivered at "ultra-high" dose rate (>100 Gy/s). For this type of irradiation, diffusion and metabolism of oxygen can be neglected and the effect of ROD on oxygen enhancement depends only on the prescribed single-fraction dose and initial oxygen concentration in the tissue. For a single-fraction dose of 10 Gy, the biological effect of

ROD is expected only in hypoxic tissues (pO2 of 2-5 mmHg) based on the expected oxygen depletion rate of 0.42 mmHg/Gy.

The second model considers the case where radiation is delivered at a "very high" dose rate, defined as 10 Gy/s or higher. Using the diffusion equation to model the replenishment of interstitial oxygen, we find that the effect of ROD diminishes as radiation pulses increase in duration. While no clear threshold is observed, no significant radioprotection is expected for pulses of radiation longer than 1-10 seconds. Additionally, fractionation of the radiation into successive pulses limited the benefits of ROD since oxygen is replenished in between pulses. Finally, assuming a single oxygenated capillary vessel, the results of our simulation suggest that only a narrow volume of tissue around the capillary would benefit from increased sparing by ROD during FLASH irradiation. The exact location of these tissues being spared depends on the oxygen tension in the capillary and the oxygen metabolism of the tissue.

Initial findings of normal tissue sparing by FLASH irradiation were received with great interest given the potential for reducing normal tissue toxicities. However, the mechanism behind this effect remains subject to debate. Several biological explanations have been advanced, including sparing of circulating immune cells due to the rapid treatment time, changes in chromatin remodeling, DNA damage repair kinetics, and inflammatory/anti-inflammatory cell signaling (Durante *et al.*, 2018). Recently, the production of organic hydroperoxides and peroxyl radicals derived through lipid peroxidation chain reactions and higher levels of redox-active iron have been advanced as possible explanations for the differential sparing of normal tissue by FLASH radiation (Spitz *et al.*, 2019).

This computational study focuses on the putative role of ROD as an OER-modifying factor. A number of studies have shown increased cell survival *in vitro* when cells were irradiated at ultra-high dose rate under hypoxic conditions, corroborating our simulation results (Berry and Stedeford, 1972; Epp *et al.*, 1972; Nias *et al.*, 1969). However, ROD may not explain the effect of FLASH on normal tissue cells, most of which are well oxygenated. Our simulations in perfused tissues show that cells must be sufficiently far from capillaries to experience a survival benefit from ROD. This finding led us to hypothesize that tissue sparing by FLASH irradiation may be driven by normal tissue stem cells, which may reside in hypoxic niches away from the vasculature.

The role of normal tissue stem cells in recovery from radiation induced damage was reported for cognitive (Acharya *et al.*, 2009), gastrointestinal (Potten *et al.*, 1997) and bone marrow toxicity (Mauch *et al.*, 1995). Direct measurements have demonstrated that such stem cell niches in bone marrow may exist in a hypoxic microenviroment with the lowest local p0$_2$ of approximately 9.9 mm Hg or 1.3 %) reported (Spencer *et al.*, 2014). Hypoxic microenvironments have also been reported for mesenchymal stem cell niches ( 2-8 % pO$_2$); neural stem cell niches (<1-8 % pO$_2$); and in additional studies for hematopoietic stem cell niches (1-6 % pO$_2$) (Mohyeldin *et al.*, 2010; Liu *et al.*, 2017). Stem cell niches have also been described in the colon (Samuelson, 2018), lung (Donne *et al.*, 2015) and kidney (Oliver *et al.*, 2004), although the oxygen concentrations of these stem cell niches were not reported.

From our model, the oxygen tension in these hypoxic stem cell niches overlaps with the range where one would expect a significant radioprotectant effect from additional oxygen depletion associated with FLASH. We therefore hypothesize that the observed effects on normal tissue toxicity of FLASH versus conventional irradiation may be mediated by the sparing of stem cells residing in hypoxic niches in normal tissues.

The existence of normal tissue stem cells in a hypoxic niche and the impact of clinically applied modifiers of tumor hypoxia may offer at least partial explanations for previously observed clinical and translational research findings. These considerations are based on the following assumptions: (a) the hypoxic normal tissue stem cells are relatively radioresistant; (b) when "activated" (e.g., by tissue damage requiring repair), they become more oxygenated and therefore more radiosensitive; and (c) therapeutic gain by modification of oxygenation is dependent on the relative effect of the modification on tumor hypoxic cells compared with their effect on the hypoxic normal tissue stem cell fraction: more oxygenation of the tumor (or destruction of hypoxic tumor cells) than on normal tissue stem cells would be expected to result in better tumor control; while if modification was more effective in shifting normal tissue hypoxia stem cells to oxygenated ones, then no improvement in tumor control would be expected—and perhaps an increased in normal tissue toxicity would be seen.

We consider three separate clinical studies for potential illustration of such effects:

*(i) Carbogen breathing during radiation therapy.* A trial in patients with esophageal cancer involved treatment with high dose rate intraluminal brachytherapy during the administration of carbogen. Excessive acute mucosal normal tissue toxicity was observed. This is consistent with the carbogen increasing the oxygenation of the normal tissue and "driving" the hypoxic stem cells to a more oxygenated state and therefore increasing their radiosensitivity (Hoskin *et al.*, 1996).

*(ii) Tirapazamine (TPZ) chemoradiotherapy in head and neck (H&N) cancers.* TPZ is a hypoxic cell toxin; it is a bioreductive alkylating agent that is selectively activated and toxic in hypoxic cells. An initial phase II trial in patients with advance H&N cancers (TROG 98.02) showed more febrile neutropenia and grade 3 or 4 late mucous membrane toxicity with TPZ/cisplatin/RT compared to patients treated with cisplatin/RT alone (Rischin *et al.*, 2005). This finding is consistent with killing of hypoxic normal tissue stem cells leading to worse radiation toxicity.

*(iii) Hyperbolic oxygen.* Hyperbaric oxygen has long been used in the treatment of chronic tissue damage caused by radiation. It has been shown to increase stem cell proliferation (Peña-Villalobos *et al.*, 2018). This raises concern that hyperbaric oxygen might result in radiosensitization of normal tissue stem cells and contribute to increased normal tissue toxicity.

The model presented in this article relies on a few key assumptions. First, we assume a linear model for radiation cell survival. This model is approximate since most cell types display a shoulder that is best modeled using a linear-quadratic (LQ) model parameterized by $\alpha$ and $\beta$. A linear model is accurate to describe cell survival beyond the initial shoulder. The use of a linear model is not a requirement of our approach and in fact any radiosensitivity model can be used in Equation (3). A key methodological challenge, however, is the need for robust radiosensitivity data that covers a wide range of oxygen tensions. The seminal work of Alper and Howard-Flanders resulted in a simple analytical model describing the linear radiosensitivity of cells as a function of oxygen tension (Alper and Howard-Flanders, 1956b). Since then, various studies have attempted to estimate OER according to the LQ model (Carlson *et al.*, 2006; Skarsgard and Harrison, 1991; Wouters and Brown, 1997; Nahum *et al.*, 2003). However, several technical challenges have emerged from these studies. When OER is defined as a ratio of doses, the ratio becomes a function of the cell killing endpoint. Alternatively, OER can be defined in terms of radiosensitivity but then two independent ratios must be introduced, $OER_\alpha$ and $OER_\beta$. It is also unclear how these factors vary as a function of oxygen tension, as most studies have only considered the extreme cases of anoxia and nornoxia. For these reasons, a linear radiosensitivity model was assumed in our simulations.

Second, our model assumes that ROD and DNA damage fixation occur instantaneously once ionizing energy is deposited into the cell. In fact, these processes occur on the time scale of microseconds (Colliaux et al., 2015) and therefore the proposed model may not be entirely accurate for radiation pulses significantly shorter than 100 μs. An improved model may include a physicochemical module that simulates chemical reactions following the formation of water radiolysis products.

Third, we assumed that the rate of ROD in tissue was similar to the rate reported in water and cell culture medium. Recently, it has been hypothesized that the rate of ROD may be 4 times higher in brain tissue than in water due to lipid peroxidation chain reactions and reactions driven by redox-active iron (Spitz et al., 2019). The computation model presented in this article can be used to estimate oxygen depletion in brain tissues based on the higher rate of ROD proposed by Spitz et al. As shown in Figure 7, a higher ROD rate of 1.8 mmHg/Gy (equivalent to 2.5 µM/Gy) would be sufficient to protect cells from a single-fraction dose of 10 Gy for oxygen tension up to 20 mmHg. Given the range of oxygen tension in normal peritumoral brain tissue (Collingridge *et al.*, 1999), radiation protection of normal tissues by FLASH could occur without invoking special hypoxic niches. The rate of ROD in various tissues must be measured experimentally before a definite conclusion can be reached regarding the role of stem cell niches in FLASH irradiation.

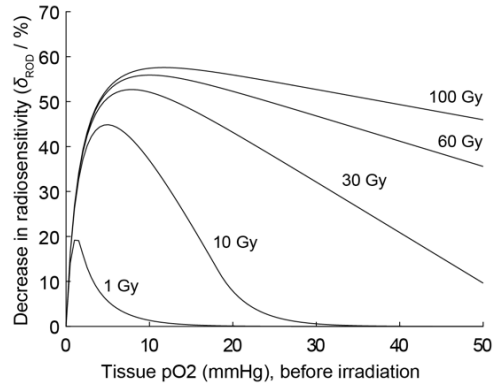

**Figure 7.** Relative decrease in radiosensitivity due to ROD assuming ultra-high-dose-rate radiation (>100 Gy/s) and high ROD rate $L = 1.8$ mmHg/Gy as proposed by Spitz *et al.* (2019).

The detrimental effect of tumor hypoxia on the control of murine and human tumors by ionizing radiation has been widely studied (Vordermark and Horsman, 2016). The confounding effect of acute hypoxia on tumor control has been less widely investigated. One study found that the removal of approximately 30 % of blood immediately prior to irradiation of tumor-bearing mice caused a high degree of radioresistance equivalent to an increase in hypoxic fractions by factors of 10-30. Measurement of $^{14}$C-misondiazole binding to these tumors after acute hypoxia indicated changes in the number of hypoxic cells corresponding to the survival endpoint (Hirst and Wood, 1987).

Studies in patients using Eppendorf electrodes to map tumors have shown significant heterogeneity in oxygen tension including regions of hypoxia (Wong et al., 1997). Variable levels of hypoxia have been demonstrated in a wide variety of human tumors including carcinoma of the cervix, squamous cell carcinoma of the head and neck, soft tissue sarcoma, as well as pancreatic, prostate, and breast cancers. The extent of intratumoral hypoxia has been associated with worse survival in cervical cancers treated with radiation. For example, disease-free survival rate was significantly worse for cancers where the percentage of pO$_2$ readings < 5 mm Hg (HP$_5$) was greater than 50 % compared to those tumors with HP$_5$ < 50 % (Fyles *et al.*, 1998). Similarly, survival rates were lower in cervical cancer patients with median pO$_2$ < 10 mm Hg compared with those with median pO$_2$ > 10 mm Hg (Fyles *et al.*, 1998) and 3-year local control rates were worse for patients with tumors with pO$_2$ < 20 mm Hg compared with those with tumors with pO$_2$ > 20 mm Hg (Suzuki *et al.*, 2006). For patients with head and neck carcinomas, worse disease-free survival was noted for patients with median tumor pO$_2$ < 10 mm Hg compared to those with pO$_2$ > 10 mm Hg (Brizel *et al.*, 1997). In addition, more hypoxic soft tissue sarcomas (median pO$_2$ < 10 mm Hg) had higher likelihood of distant metastases (Brizel *et al.*, 1996) and, in another study, hypoxic soft tissue sarcomas had a poorer disease-specific and overall survival at 5 years (Nordsmark *et al.*, 2001). Oxygen tension distributions were also sufficient to explain the local response of human breast tumors treated with radiation (Okunieff *et al.*, 1993).

In light of this association between tumor hypoxia and radioresistance, a transient increase in the extent of hypoxia due to ROD could render chronically hypoxic tumors more resistant to FLASH irradiation. This potential effect of FLASH on hypoxic tumors should be further investigated.

## Conclusions

We presented two models of ROD during FLASH irradiation, a simplified one which is valid for ultra-high dose rate and a more detailed one that includes oxygen diffusion and replenishment during radiation. Neither of these models provides a definitive answer as to what mechanism may be responsible for the FLASH effect, but they can be used to generate hypotheses that can be tested experimentally. For instance, the model predicts that the FLASH effect should gradually disappear as the radiation pulse duration is increased from <1s to 10 s. Second, fractionation of the dose as a series of pulses should be less efficient than the delivery of the same dose as a single pulse. Finally, changes in capillary oxygen tension (increase or decrease) should result in a decrease of the FLASH effect. This last prediction could be tested using a hyperbaric chamber or through carbogen or nitrogen breathing.